\documentclass[aps,twocolumn,pra,%
preprintnumbers,amsmath,amssymb,superscriptaddress]{revtex4}
\usepackage{graphicx}
\usepackage{color}
\newcommand{\bea}{\begin{eqnarray}}
\newcommand{\eea}{\end{eqnarray}}

\newcommand{\mc}[1]{\mathcal{#1}}

\def\slash#1{\not\!\!#1}

\usepackage{ulem} %

\begin{document}

\preprint{OU-HET-891}
\preprint{RIKEN-STAMP-25}

\title{Comments on 
SUSY Exact Action 
in 3D Supergravity}

\author{Norihiro {Iizuka}}\email[]{iizuka@phys.sci.osaka-u.ac.jp} 
\affiliation{{\it Department of Physics, Osaka University, Toyonaka, Osaka 560-0043, JAPAN}}

\author{Akinori {Tanaka}}
\email[]{akinori.tanaka@riken.jp}
\affiliation{{\it 
Interdisciplinary Theoretical Science Research Group, 
RIKEN, Wako 351-0198, JAPAN}}

\begin{abstract}
We consider 2+1 dimensional off-shell $\mc{N}=1$ pure supergravity that is constructed from graviton, gravitino and auxiliary field.
We show that the $R^2$ supersymmetric invariant and $R_{\mu \nu}^2$ supersymmetric invariant are expressed as local supersymmetric exact terms 
up to 
mass terms for gravitino.
In both cases, the mass parameter is proportional to the off-shell supersymmetric cosmological constant. 
\end{abstract}

\maketitle



\noindent

\section{Introduction}
Calculating quantum gravity partition function in a certain reasonable way is one of the most 
important {and fundamental} questions 
in theoretical physics.
Even in the conventional quantum field theory with spin 0, 1/2, 1 fields, the exact computation is extremely difficult in many cases, and we are often tempted to use perturbative analysis. 
However, if there are some supersymmetries, one can utilize these symmetries to reduce the path integral to the finite dimensional matrix models \cite{Pestun:2007rz, Kapustin:2009kz}.
In this procedure, the existence of supersymmetric ``exact" Lagrangian is extremely important because adding such term into the path integral weight does not change the final result and the WKB computation turns out to be exact by taking its coupling constant to be infinite (or zero). 

By applying this technique to the gravity path integral, we would like to make the gravity path integral well-defined.    
In \cite{Iizuka:2015jma}, the authors considered such possibility in terms of supersymmetric Chern-Simons formulation of the three-dimensional gravity.
In this notes, we discuss another possibility: localization computation with local supersymmetry.
We will focus on 2+1 dimensional $\mc{N}=1$ supergravity, and we start with 
reviewing some known facts on the theory. 

\section{3D 
$\mc{N}=1$ off-shell supergravity} 
We focus on the Lorentz signature
$
\eta^{ab} = \text{diag} (-1, +1, +1),
$
where the alphabet runs for local Lorentz indices $a,b = 0,1,2$.
The fundamental degrees of freedom are graviton $e{_\mu}{^a}$, gravitino: $\psi{_\mu}$, real auxiliary field, $S$.
Local supersymmetry is defined by an arbitrary Majorana spinor parameter $\epsilon$ which depends on the coordinates as follows \cite{Andringa:2009yc}:
\bea
\delta e{_\mu}{^a} 
&=& \frac{1}{2} (\bar{\epsilon} \gamma^a \psi_\mu)
\,, \quad
\delta \psi_\mu 
\,=\, D_\mu (\hat{\omega}) \epsilon + \frac{1}{2} S \gamma_\mu \epsilon
\,,
\\
\delta S 
&=& \frac{1}{4} (\bar{\epsilon} \gamma^{\mu \nu} \psi_{\mu \nu} (\hat{\omega})
- \frac{1}{4} (\bar{\epsilon} \gamma^\mu \psi_\mu ) S \,,
\eea
where the covariant derivative is defined by $D_\mu (\omega) = \partial_\mu + \frac{1}{4}\omega^{ab}_\mu \gamma_{ab}$ and the hatted spin connection contains the contribution from torsion. 
And $\psi_{\mu \nu} (\omega) = \frac{1}{2} (D_\mu(\omega) \psi_\nu - D_\nu(\omega) \psi_\mu  )$.
See \cite{Andringa:2009yc} for more details.

Under these transformations, the following Lagrangians are invariant up to total derivative term.
\bea
L_{EH}
&=&
e \, 
(R - \bar{\psi}_\mu \gamma^{\mu \nu \rho} D_\nu (\hat{\omega}) \psi_\rho
- 2 S^2)
\,,
\\
L_{C}
&=&
e \, 
(S + \frac{1}{8} \bar{\psi}_\mu \gamma^{\mu \nu} \psi_\nu) \,. 
\label{cosm}
\eea
The first one is the usual Einstein-Hilbert term.
The second one corresponds to the cosmological constant term.
Just by integrating out the auxiliary field, it generates the usual negative cosmological constant term, 
and the resultant Lagrangian turns out to be so-called $\mc{N}=(1,0)$ AdS-supergravity.
In addition to that, one can find supersymmetric gravitational CS term, but we omit them for simplicity.
Other supersymmetric terms can be found in \cite{Andringa:2009yc} as follows.
\begin{align}
&L_{R_{\mu \nu}^2}
\, = \,
- \frac{1}{4} e R^{\mu \nu ab} (\Omega^+) R_{\mu \nu ab} (\Omega^+)
\notag \\
& \,
- 2 e \bar{\psi}_{ab} (\Omega^-) \gamma^\mu D_\mu \psi^{ab} (\Omega^-)
+ \frac{1}{2} e R_{\mu \nu ab} (\Omega^+) \bar{\psi}_\rho \gamma^{\mu \nu} \gamma^\rho \psi^{ab}(\Omega^-)
\notag \\
& \quad 
+e S \bar{\psi}_{ab} (\Omega^-) \psi^{ab} (\Omega^-)
- \frac{1}{2} e \bar{\psi}^{ab} (\Omega^-) \psi_{ab} (\Omega^- ) \bar{\psi}_\mu \psi^\mu
\notag \\
& \qquad 
+\frac{1}{8} e \bar{\psi}^{ab} (\Omega^-) \psi_{ab} (\Omega^- ) \psi_\mu \gamma^{\mu \nu} \psi_\nu,
\end{align}
where $\Omega^{\pm ab}_\mu = \hat{\omega}{_\mu}{^{ab}} \pm S \varepsilon{_\mu}{^{ab}} $ and $(\Omega^\pm)$ means that the corresponding object is defined by the covariant derivative with respect to $\Omega^\pm$.
This is $R_{\mu \nu}^2$ type supersymmetric Lagrangian. A nice property is that $L_{R_{\mu \nu}^2}$ can be represented as the supersymmetric Yang-Mills action by considering the pair of indices $ab$ as the gauge index and regarding the gauge field $A_\mu^{I} = \Omega{^+_\mu}{^{ab}}$ and the gaugino $\chi^I = \psi^{ab} (\Omega^-)$:
\begin{align}
&
L_{SYM}
\notag \\
&=
- \frac{1}{4} e F^{\mu \nu I} F{_{\mu \nu}}{^I}
- 2 e \bar{\chi}^I \gamma^\mu (D_\mu \chi)^I
+ \frac{1}{2} e F{_{\mu \nu}}{^I} \bar{\psi}_\rho \gamma^{\mu \nu} \gamma^\rho \chi^I
\notag \\
&
+ e S \bar{\chi}^I \chi^I
- \frac{1}{2} e \bar{\chi}^I \chi^I \bar{\psi}_\mu \psi^\mu
+ \frac{1}{8} e \bar{\chi}^I \chi^I \bar{\psi}_\mu \gamma^{\mu \nu} \psi_\nu. 
\notag \\
&=
L_{R_{\mu \nu}^2}
\label{sym}
\end{align}
where we use the identifications $A_\mu^{I} = \Omega{^+_\mu}{^{ab}}$ and $\chi^I = \psi^{ab} (\Omega^-)$ in the final equality. 

In addition, one can also find the following $R^2$ type supersymmetric term, 
\begin{align}
&L_{R^2} 
\notag \\
&
=\frac{1}{16} e \hat{R}^2 (\Omega^+)
+ \frac{1}{4} e \bar{\psi}_{\mu \nu} (\Omega^-) \gamma^{\mu \nu} \slash{D} \psi_{\rho \sigma} (\Omega^-) - e \partial^\mu S \partial_\mu S 
\notag \\
&
- \frac{1}{8} e S \bar{\psi}_{\mu \nu} (\Omega^-) \gamma^{\mu \nu} \gamma^{\rho \sigma} \psi_{\rho \sigma} (\Omega^-)
+ \frac{1}{2} e \bar{\psi}_\mu \gamma^\nu \gamma^\mu \partial_\nu S \gamma^{\rho \sigma} \psi_{\rho \sigma} (\Omega^-)
\notag \\
&
- \frac{1}{32} e \bar{\psi}_{\mu \nu} (\Omega^-) \gamma^{\mu \nu} \gamma^{\rho \sigma} \psi_{\rho \sigma} (\Omega^-) \bar{\psi}_\lambda \psi^\lambda
\notag \\
&
+\frac{1}{64} e \bar{\psi}_{\mu \nu} (\Omega^-) \gamma^{\mu \nu} \gamma^{\rho \sigma} \psi_{\rho \sigma} (\Omega^-) \bar{\psi}_\lambda \gamma^{\lambda \tau} \psi_\tau,
\end{align}
where the hatted curvature is defined by
\begin{align}
&\hat{R} (\Omega^+)
=
R(\hat{\omega})
+ 6 S^2 \notag \\
& \qquad  \qquad 
+ 2 \bar{\psi}_\mu \gamma_\nu \psi^{\mu \nu} (\Omega^-)
+\frac{1}{2} S \bar{\psi}_\mu \gamma^{\mu \nu} \psi_\nu
.
\end{align}
Similarly one can regard this $L_{R^2}$ matter Lagrangian 
as follows: 
\begin{align}
&L_{\text{matter}} 
\notag \\
&=
-e \partial^\mu \phi \partial_\mu \phi
- \frac{1}{4} e \bar{\lambda} \gamma^\mu D_\mu \lambda
+ \frac{1}{16} e f^2
+ \frac{1}{8} e S \bar{\lambda} \lambda
\notag \\
&
+ \frac{1}{2} e \bar{\psi}_\mu \gamma^\nu \gamma^\mu \partial_\nu \phi \lambda
+ \frac{1}{32} e \bar{\lambda} \lambda \bar{\psi}_\mu \psi^\mu
- \frac{1}{64} e \bar{\lambda} \lambda \bar{\psi}_\mu \gamma^{\mu \nu} \psi_\nu.
\notag \\
&=
L_{R^2}
\label{mat}
\end{align}
where we use the identifications for the scalar 
$\phi = S$, for the spinor $\lambda = \gamma^{\mu \nu} \psi_{\mu \nu} (\Omega^-)$, and for the auxiliary scalar 
$f = \hat{R}(\Omega^\pm)$ 
%
in the final equality.

\section{SUSY Exact terms}  
 
%
For the localization calculation, the most important feature is the following point:
To obtain the partition function $Z = \lim_{t\to0} Z(t)$, we define $Z(t)$ as 
\begin{align}
Z(t)=
\int 
\mathcal{D}e{_\mu}{^a} 
\mathcal{D} \psi_\mu
\mathcal{D} S
\
e^{iS + it \delta V} \,,
\label{ptn}
\end{align}
and furthermore this $Z(t)$ does not depend on the parameter $t$.
Then we can take $t \to \infty$ limit to conduct the computation, and in this limit, all the contributions of the path integral are localized on the field configurations which satisfy $\delta V =0$ \footnote{
Here, we consider the partition function with Lorentz signature. So it is plausible to take the weight as $e^{iS}$.
In the $t\to \infty$ limit, the configuration which satisfy $\delta V=0$  is dominant thanks to the Riemann-Lebesgue lemma.
}.
In quantum field theory, 
this technique achieved great successes \cite{Pestun:2007rz, Kapustin:2009kz} and uncovered structures of the interacting supersymmetric field theories in various dimensions. 
The necessary ingredients for this $t$-independence are 1: off-shell supersymmetry $\delta$, 2: supersymmetric invariant action $S=\int L$ and 3: supersymmetric exact action $\delta V$ where $V$ is a certain functional of the fields, which satisfy $\delta^2 V =0$.
Naively, we expect that its analog to the supergravity provides us an unknown structures of quantum gravity.
In this notes, we try to do it. 

In order to apply the above localization argument, the missing piece is the supersymmetric exact action $\delta V$, and 
we find that the following actions are candidates for the appropriate actions $\delta V$;  
\begin{align}
L_{R_{\mu\nu}^2 + \text{cosm}}
&=
- \frac{1}{8} L_{R_{\mu \nu}^2}
- \frac{1}{4} L_C \bar{\psi}^{ab} (\Omega^-) \psi_{ab}(\Omega^-)
\label{Rmunu2} \\
\Big( &=- \frac{1}{8} L_{SYM}
- \frac{1}{4} L_C \bar{\chi} \chi
\Big) \,,
\notag
\\
L_{R^2 +\text{cosm}}
&=
L_{R^2}
+ \frac{1}{4} L_C \bar{\psi}_{ab} (\Omega^-) \gamma^{ab} \gamma^{cd} \psi_{cd} (\Omega^-)
\label{R2}
 \\
\Big( &=
L_{\text{matter}}
- \frac{1}{4} L_C \bar{\lambda} \lambda
\Big)
\notag
\,,
\end{align}
where $L_C$ is the supersymmetric cosmological constant given in \eqref{cosm}.
In fact, one can verify the following relations;  
\begin{align}
&\delta (e[\bar{\chi} \delta \chi] ) = (\bar{\epsilon} \epsilon) L_{R_{\mu\nu}^2 + \text{cosm}}  
\label{SE1}
\\
&\delta (e[\bar{\lambda} \delta \lambda] ) = (\bar{\epsilon} \epsilon) L_{R^2 + \text{cosm}} 
\label{SE2}
\end{align}
These relations show that above $L_{R_{\mu\nu}^2 + \text{cosm}}$ and $L_{R^2 +\text{cosm}}$ are SUSY exact terms.  
Of course, the Lagrangians $L_{R_{\mu\nu}^2}$ and $L_{R^2}$ preserves supersymmetry.
However, in each case \eqref{Rmunu2} or \eqref{R2}, one has a \textit{mass term} for the fermion, and it is a typical {\it supersymmetry breaking term}, where the supersymmetry breaking is given by the supersymmetric cosmological constant term $L_C$. 

One might wonder why these SUSY exact actions are not SUSY invariant.
The reason is as follows.
In rigid limit, we have $\delta^2 = 0$ in field theoretical sense, and one can show SUSY invariance just by acting additional $\delta$ to \eqref{SE1} or \eqref{SE2}. However if we do not take a 
 rigid limit $(\psi_\mu =0)$, then we have $\delta^2 \neq 0$. As a result,  \eqref{SE1} and \eqref{SE2} are {\it not} SUSY invariant, even though they are SUSY exact.

\section{Naive attempt toward gravity localization}

Let us discuss the localization argument on supergravity based on the results in previous section.
As explained above, the only embarrassing term is the mass term for graviton $\psi_\mu$, or equivalently $\chi$ or $\lambda$ in \eqref{Rmunu2} or \eqref{R2}, which prevails the SUSY invariance of SUSY exact term \eqref{SE1} and \eqref{SE2}.   
To overcome the problem, here we try to eliminate it just by inserting the delta function $\delta(L_c)$ to the path integral in \eqref{ptn};  
\begin{align}
Z(t) = 
\int 
\mathcal{D}e{_\mu}{^a} 
\mathcal{D} \psi_\mu
\mathcal{D} S
\
\delta (L_C)
e^{i\int d^3 x L_{EH} + i \int d^3 x L_C + it \delta V}
,
\label{ourptn}
\end{align}
where we take $S=\int (L_{EH} + L_C)$, and $\delta V$ is the one \eqref{Rmunu2} or \eqref{R2}.
It might look strange, but since the delta function can be written 
by introducing auxiliary field $\varphi$ as an integral formula,  
\begin{align}
\delta (L_C)
=
\int \mathcal{D} \varphi
\
e^{i \int d^3 x L_C \cdot \varphi},
\end{align}
we can rewrite \eqref{ourptn} as
\begin{align}
Z(t) = 
\int
\mathcal{D}e{_\mu}{^a} 
\mathcal{D} \psi_\mu
\mathcal{D} S
\mathcal{D} \varphi
\
e^{i\int d^3 x L_{EH} + i \int d^3 x L_C (1 + \varphi) + it \delta V}.
\end{align}
If the supersymmetric invariance for the deformed cosmological constant term and for the path integral measure are achieved, 
then this $Z(t)$ becomes $t$-independent, and we can utilize the localization technique by taking $t \to \infty$ limit. 
For that purpose, we require
\begin{align}
\delta (\varphi L_c)  =0.
\label{varLccondition}
\end{align} 
If the SUSY variation of the cosmological constant term is total divergence, say, $\delta L_C = \nabla_\mu J^\mu$, then 
\eqref{varLccondition} implies that $\delta \varphi$ should be defined linear with respect to $\varphi$ such as  
\begin{align}
\delta \varphi = - \frac{   \nabla_\mu J^\mu}{L_C} \varphi
\,.
\end{align}
However this induces quantum anomaly, {\it i.e.} Jacobian for supersymmetry variation of $\varphi$  is not one.
In order to apply the conventional supersymmetric localization technique, the Jacobian for supersymmetry variation 
should vanish, therefore this naive method does not work, unfortunately.

\section{Conclusion, discussion, and future work}
In this notes, we discussed a possibility for the application of localization technique to the quantum gravity path integral.
We tried to conduct direct gravity path integral by constructing SUSY exact terms in 3D supergravity. Although SUSY exact terms are constructed, naive procedure for localization calculation fails. 
Our main discovery is that the $R^2_{\mu \nu}$ supersymmetric invariant, $L_{R_{\mu\nu}^2}$, and $R^2$ supersymmetric invariant, $L_{R^2}$, can be represented as SUSY \textit{exact} terms {\it up to gravitino mass terms}, which break supersymmetric invariance and its breaking is given by the 
supersymmetric cosmological constant term $L_C$. This prevents us from applying a naive localization technique to supergravity within these setups \footnote{
However, as commented in \cite{Andringa:2009yc}, the bosonic part of the equation $L_{R^2}=0$ 
is exactly equivalent to the integrability condition for the Killing spinor equation
which is equivalent to the condition
$
\delta \psi_\mu =0.
$
This fact might illuminate the possibility of localization calculous with $L_{R^2}$ action, but of course, we should overcome the graviton mass term problem in \eqref{SE2}.
}.
We would like to make some comments about our (rather negative) results.

First, let us comment on the difficulty of the gravity sector localization computation.
In our case, as one can find the algebraic structure of local SUSY $\delta$ on 3D supergravity in \cite{Achucarro:1987vz}, squared SUSY $\delta^2$ is not zero and contains SUSY $\delta$, too.
This structure is coming from the existence of gravitino, and it is absent in the rigid SUSY limit $\delta_{\text{rigid}}$ \cite{Festuccia:2011ws} which guarantees the localization computation because of the nilpotent nature $\delta^2_{\text{rigid}}=0$ in many cases.
However, the possibility for localization in supergravity is not excluded even for 3D $\mc{N}=1$ because what we 
found is 
just the relationship \eqref{Rmunu2} - 
\eqref{SE2}. 
Therefore, if one can find certain better SUSY exact terms and succeed in canceling the obstructing mass term, then it should work.

Second, the mass terms in our SUSY exact Laglangians, \eqref{Rmunu2} and \eqref{R2}, seem to be ``universal" mass terms because they are always proportional to the supersymmetric cosmological constant $L_C$ in \eqref{cosm}. We have no a priori reason to get such supersymmetric coefficient as the mass parameter, but there might exist certain deep reason which could be related to the algebraic structure on supergravity.

Third, it may be good to consider the same problem with extended local supersymmetries, $\mc{N} \geq 2$.
For example, we can find off-shell formulation of $\mc{N}=2$ supergravity in \cite{Kuzenko:2013uya}.
In 3D, conventional field theoretical localization computation is available only for $\mc{N} \geq 2$, therefore situation there could be better.

It will be  also interesting to consider the analog of our argument with \textit{Euclidean} supergravity.
(For relevant works on 3D Euclidean {\it pure} gravity with negative cosmological constant, see for example, 
\cite{Witten:2007kt, Maloney:2007ud, 
Iizuka:2015jma}.) 
Crucial difference is that in Euclidean signature,  modular invariance is strong enough to determine (some of) non-perturbative effects. It would be great if we can derive the summation over Modular group discussed in \cite{Dijkgraaf:2000fq, Manschot:2007zb, Manschot:2007ha, Maloney:2007ud} in direct supergravity localization calculation without relying on the power of modular invariance. This should be done along the line of \cite{Iizuka:2015jma}, where the sum over modular group appears naturally as the sum over all of the localization locus, ${\cal{F}}_{\mu\nu} = 0$, which are solutions of all the complex Einstein equation. 



Before we end, let us discuss the physical meaning of conducting gravity path integral, $Z = \int [{\cal{D}} g_{\mu\nu}] e^{i S[g_{\mu\nu}]}$. Even if we succeed in conducting the metric path integral $[{\cal{D}} g_{\mu\nu}]$ exactly, 
whether it gives an {\it exact} partition function for quantum gravity or not,  
depends on whether the metric $g_{\mu\nu}$ is a fundamental degree of freedom in quantum gravity. 
We have learned from holography that  
bulk gravity is an effective theory, which is valid and emerging typically in the large $N$ limit of QCD-like $SU(N)$ gauge theory as a dual effective description.   
Furthermore, a metric, which is dual to gauge-singlet stress-tensor,
is a dominating 
degree of freedom 
only in low temperature phase \cite{Witten:1998qj, Witten:1998zw}.  
%
In fact, in high temperature phase, rather than metric, 
black hole microstates are the dominating 
degrees of freedom \footnote{These can easily seen   
from comparison of the $N$-dependence of the entropy between thermal-gas/black holes in bulk and 
confinement/deconfinement phase in the boundary 
\cite{Witten:1998qj, Witten:1998zw}.}.    
Given these, how much is the bulk metric path integral meaningful calculation? 

To answer this, the analogy to QCD helps;  
gravity in low temperature phase is like chiral Lagrangian in QCD, where the dynamical degrees of freedom are pion field $\pi$'s,  instead of quarks and gluons. Then  
conducting 
gravity path integral $ \int [{\cal{D}} g_{\mu\nu}] e^{i S[g_{\mu\nu}]}$ 
corresponds to 
conducting 
pion field path integral $\int [{\cal{D}} \pi] e^{i S_{chiral}}$ in the chiral lagrangian. 
%
Of course we know the fundamental theory behind chiral lagrangian is QCD, and the {\it exact} answer for the partition function for QCD can be obtained only after by conducting the path integral for quark-gluon fields, rather than pion fields. Pion field path integral of the chiral lagrangian never gives the right answer for QCD, due to its lack of quark and gluon degrees of freedom which are dominating in high temperature phase  
\footnote{If we pay attention to physics only in the confinement phase, 
namely, only low temperature phase and neglect all non-perturbative effects, then the results of pion path integral of chiral lagrangian is still meaningful as an effective theory.}.  
As one cannot describe quark-gluon plasma by multi-pion fields, 
we expect that 
black hole microstates are not describable by 
multi-gravitons (see \cite{ElShowk:2011ag} for a nice overview) 
\footnote{Note also that the argument of \cite{Witten:1998qj, Witten:1998zw} works 
in bulk where we have
 space-time dimensions larger than three. Even in  
three dimensional space-time case, BTZ black hole microstates are regarded as a different primary's conformal family \cite{Witten:2007kt,  Maloney:2007ud, Iizuka:2015jma, Honda:2015hfa}.}.  
In this way, 
we expect that the naive bulk metric path integral, $Z = \int [{\cal{D}} g_{\mu\nu}] e^{i S[g_{\mu\nu}]}$, 
is not non-perturbatively-defined quantity, at least in bulk where we have space-time dimensions larger than three. (Note however in three-dimension, modular invariance of the partition function is powerful enough to determine the contributions of BTZ black hole microstates, see \cite{Witten:2007kt,  Maloney:2007ud, Iizuka:2015jma, Honda:2015hfa}.) 
To obtain an exact partition function for full quantum gravity, we have to rely on the dual non-perturbatively defined boundary theory path integral 
\footnote{It gives, at most, approximately valid and meaningful quantity only in low temperature phase, {\it i.e.}, bulk phase without black holes.}.  

However what we try to calculate in this paper is {\it not} this quantity (partition function), but rather supersymmetric index due to the fermion boundary condition. 
Then the situation is totally different: 
Index calculations in field theory quite often works to count the supersymmetric black hole microstates.  
This is because 
of supersymmetry, 
significant reduction of degrees of freedom occurs.     
Therefore, 
the SUSY index calculation from the bulk metric by conducting $\int [{\cal{D}} g_{\mu\nu}]$ is still 
meaningful even in bulk.   

\acknowledgments
%
{This work was also supported in part by 
JSPS KAKENHI Grant Number 25800143 (NI).
This work was partially supported by the RIKEN iTHES Project.
}


\appendix
{\section{Spinor notations and formulas}}
The clifford algebra is generated by the following two by two matrices.
\begin{align}
\gamma^0
=
\begin{pmatrix}
0 & -1 \\
1 & 0
\end{pmatrix},
\gamma^1
=
\begin{pmatrix}
0 & 1 \\
1 & 0
\end{pmatrix},
\gamma^2
=
\begin{pmatrix}
1 & 0 \\
0 & -1
\end{pmatrix}.
\end{align}
Charge conjugation matrix is
\begin{align}
C = i \gamma^0.
\end{align}
We always use Majorana fermion throughout this notes.
It satisfies
\begin{align}
\bar{\psi} = \psi^\dagger i \gamma^0 = \psi^T C
\end{align} 
and it is equivalent to $\psi^* = \psi$ and it means real fermion.
For Majorana fermions $\psi, \chi, \epsilon$, we have the following formulas:
\begin{align}
&\bar{\psi} \chi = \bar{\chi} \psi,
\\
&\overline{(\gamma^\mu \psi)} \chi = - \bar{\psi} \gamma^\mu \chi,
\\
& \bar{\chi} \gamma^\mu \psi = - \bar{\psi} \gamma^\mu \chi,
\\
& \bar{\chi} \gamma^\mu \gamma^\nu \psi = \bar{\psi} \gamma^\nu \gamma^\mu \chi,
\\
& \bar{\chi} \gamma^\mu \gamma^\nu \gamma^\rho \psi = - \bar{\psi} \gamma^\rho \gamma^\nu \gamma^\mu \chi,
\\
& \epsilon (\bar{\chi} \psi) + 2 (\bar{\chi} \epsilon) \psi + \gamma^\mu \epsilon (\bar{\chi} \gamma_\mu \psi) =0
\\
& (\bar{\psi} \epsilon) (\bar{\epsilon} \chi) = - \frac{1}{2} (\bar{\epsilon} \epsilon) (\bar{\psi} \chi) \label{A10}.
\end{align}
The formula in \eqref{A10} is useful in the calculation of \eqref{SE1} and \eqref{SE2}.
{\section{Proof of \eqref{SE1} }}
As commented in the main part of this notes, if we define $A_\mu^I = \Omega_\mu^{+ ab}, \chi^I = \psi^{ab} (\Omega^-)$, these redefined multiplet satisfy
\begin{align}
\delta A_\mu = - (\bar{\epsilon} \gamma_\mu \chi),
\quad
\delta \chi = \frac{1}{8} \gamma^{\mu \nu} (F_{\mu \nu} + 2 \bar{\psi}_{[\mu} \gamma_{\nu]} \chi ) \epsilon,
\end{align}
where we omit the index $I$ from now on.
Just using these SUSY transform, we calculate
\begin{align}
\delta \Big[ e (\bar{\chi} \delta \chi) \Big]
=
\delta e \cdot (\bar{\chi} \delta \chi)
+
e (\overline{\delta \chi} \delta \chi)
+
e (\bar{\chi} \delta^2 \chi)
\end{align}
as follows. We use \eqref{A10} many times.
\begin{align}
&\delta e \cdot (\bar{\chi} \delta \chi)
=
(\bar{\epsilon} \epsilon)
e \Big\{
- \frac{1}{32} (\bar{\chi} \gamma^{\mu \nu} \gamma^\rho \psi_\rho) F_{\mu \nu}
- \frac{1}{16} (\bar{\chi} \chi) (\bar{\psi}_\mu \gamma^{\mu \nu} \psi_\nu)
\notag \\ & \qquad \qquad \qquad \qquad
- \frac{1}{16} (\bar{\chi} \chi) (\bar{\psi}_\mu \psi^\mu)
\Big\}
\label{B3}
\\
&e (\overline{\delta \chi} \delta \chi)
=
(\bar{\epsilon} \epsilon) e
\Big\{
\frac{1}{32} F_{\mu \nu} F^{\mu \nu}
+ \frac{1}{8} (\bar{\psi}^\mu \gamma^\nu \chi) F_{\mu \nu}
\notag \\ & \qquad \qquad\qquad\qquad
+ \frac{1}{16} (\bar{\chi} \chi) (\bar{\psi}_\mu \psi^\mu)
+ \frac{1}{32} (\bar{\chi} \chi) (\bar{\psi}_\mu \gamma^{\mu \nu} \psi_\nu)
\Big\}
\label{B4}
\\
&e (\bar{\chi} \delta^2 \chi)
\notag \\
&=
(\bar{\epsilon} \epsilon) e
\Big\{
\big[ \frac{1}{16} (\bar{\chi} \gamma^{\rho \nu} \gamma^\mu \psi_\rho)
+\frac{1}{32} (\bar{\psi}_\rho \gamma^{\mu \nu} \gamma^\rho \chi)
+ \frac{1}{16} (\bar{\psi}^\mu \gamma^\nu \chi) \big] F_{\mu \nu}
\notag \\ & \qquad \qquad
+ \frac{1}{16} (\bar{\chi} \chi ) (\bar{\psi}_\mu \psi^\mu
- \frac{1}{64} (\bar{\chi} \chi) (\bar{\psi}_\mu \gamma^{\mu \nu} \psi_\nu)
\notag \\ & \qquad \qquad 
+ \frac{1}{4} (\bar{\chi} \gamma^\mu D_\mu \chi)
- \frac{6}{16} S (\bar{\chi} \chi) 
\Big\}.
\label{B5}
\end{align}
Summing up \eqref{B3}, \eqref{B4} and \eqref{B5}, we get the result in \eqref{SE1}.

{\section{Proof of \eqref{SE2} }}
If we define $\phi = S$, $\lambda = \gamma^{\mu \nu} \psi_{\mu \nu} (\Omega^-)$, $f=\hat{R}(\Omega^\pm)$, then these fields satisfy
\begin{align}
&\delta \phi = \frac{1}{4} \bar{\epsilon} \lambda, 
\quad 
 \delta \lambda = \gamma^\nu \epsilon [\partial_\nu \phi - \frac{1}{4} \bar{\psi}_\nu \lambda] - \frac{1}{4} \epsilon f
\notag
\\
&\delta f =
- \bar{\epsilon} \gamma^\mu [D_\mu(\hat{\omega}) \lambda  - \gamma^\nu \psi_\mu (\partial_\nu \phi - \frac{1}{4} \bar{\psi}_\nu \lambda)
+ \frac{1}{4} f \psi_\mu ]
\notag \\ & \qquad
\,\,\, + \frac{1}{2} S (\bar{\epsilon} \lambda).
\end{align}
By using these SUSY transformations, we calculate
\begin{align}
\delta \Big[ e (\bar{\lambda} \delta \lambda) \Big]
=
\delta e \cdot (\bar{\lambda} \delta \lambda)
+
e (\overline{\delta \lambda} \delta \lambda)
+
e (\bar{\lambda} \delta^2 \lambda) \,,
\end{align}
and each term is given by as follows: 
\\

%
\bea
&& \delta e \cdot (\bar{\lambda} \delta \lambda)
=
(\bar{\epsilon} \epsilon)
 e
 \Big\{ 
- \frac{1}{4} (\bar{\psi}_\mu \gamma^\mu \gamma^\nu \lambda) \partial_\nu \phi
- \frac{1}{32} (\bar{\lambda} \lambda) (\bar{\psi}_\mu \psi^\mu)
\notag \\ && \qquad \qquad \quad 
- \frac{1}{32} (\bar{\lambda} \lambda) (\bar{\psi}_\mu \gamma^{\mu \nu} \psi_\nu)
+ \frac{1}{16} f (\bar{\lambda} \gamma^\mu \psi_\mu)
\Big\} \,, 
\label{C3}
\\
&& \,\,\,\,e (\overline{\delta \lambda} \delta \lambda)
=
(\bar{\epsilon} \epsilon)
 e
 \Big\{ 
 - \partial_\mu \phi \partial^\mu \phi
 + \frac{1}{2} (\bar{\psi}^\mu \lambda) \partial_\mu \phi
  \notag \\ && \qquad \qquad \qquad \qquad
 + \frac{1}{32} (\bar{\lambda} \lambda) (\bar{\psi}_\mu \psi^\mu )
 + \frac{1}{16} f^2
 \Big\} \,, \quad 
 \label{C4} 
\\
&& \,\,\,\, e (\bar{\lambda} \delta^2 \lambda)
=
(\bar{\epsilon} \epsilon)
 e
 \Big\{ 
 \frac{1}{4} (\bar{\psi}_\mu \gamma^\nu \gamma^\mu \lambda) \partial_\nu \phi
 + \frac{1}{32} (\bar{\lambda} \lambda) (\bar{\psi}_\mu \psi^\mu)
    \notag \\ && \qquad \qquad 
 - \frac{1}{32} (\bar{\lambda} \lambda) (\bar{\psi}_\mu \gamma^{\mu \nu} \psi_\nu)
 - \frac{1}{8} S (\bar{\lambda} \lambda)
 - \frac{1}{4} (\bar{\lambda} \gamma^\mu D_\mu \lambda)
    \notag \\ && \qquad \qquad 
 +\frac{1}{64} (\bar{\lambda} \lambda) (\bar{\psi}_\mu \gamma^{\mu \nu} \psi_\nu)- \frac{1}{16} f (\bar{\lambda} \gamma^\mu \psi_\mu)
 \Big\} \label{C5}.
\eea
Combining \eqref{C3}, \eqref{C4} and \eqref{C5}, we arrive at \eqref{SE2}.


\end{document}